\documentclass[tikz,12pt,a4paper]{article}
\usepackage{amsmath, amssymb, amsfonts}
\usepackage{bm}
\usepackage{tikz}
\usepackage{graphicx}

\title{Entropy-Time Geodesics as a Universal Framework for Transport and Transition Phenomena}
\author{Sami Lakka}
\author{Sami Lakka\thanks{Corresponding author: sami.lakka@gmail.com} \\
	LAKKA Health \\
	Lempäälä, 37500, Finland}

\date{\today}

\begin{document}
	\maketitle

\begin{abstract}
	
	We develop a geometric framework for irreversible transport phenomena in which macroscopic evolution equations arise from the combined structure of a thermodynamic state metric and an Onsager-based dissipation metric. The construction begins by defining a pseudo-Riemannian manifold from the Hessian of an appropriate thermodynamic potential. When the enthalpy is used and written in variables $(S,P)$, the resulting metric possesses a Lorentzian-type signature: entropy acts as a time-like coordinate, while pressure forms a spatial-like coordinate associated with mechanical response. Local irreversible dynamics are incorporated through the inverse Onsager matrix, which defines a positive-definite dissipation metric on the space of fluxes and gradients. A thermodynamic action integrating these two geometric layers yields geodesic evolution equations. For a Newtonian fluid with constant viscosity, the resulting Euler--Lagrange equations reproduce the incompressible Navier--Stokes equations without requiring an externally imposed constitutive closure.
	
	Within this framework, turbulence scaling emerges from competition between inertial curvature and dissipation metric stiffness. The Kolmogorov length scale appears as a minimum geometric resolution length where these contributions balance, providing a geometric interpretation of energy cascade termination and dissipation onset. Finite-time singularities in the classical PDE formulation correspond to curvature divergences in the transport geometry; however, the thermodynamic proper time diverges in such limits, suggesting that blow-up is dynamically suppressed in single-phase continua. Breakdown occurs only when the thermodynamic metric itself becomes degenerate, such as at cavitation or critical points where compressibility or heat capacity diverge, corresponding to a change in manifold topology.
	
	Although derived explicitly for fluid flow, the framework is general: by choosing different thermodynamic potentials and Onsager matrices, the same geometric formulation applies to heat conduction, diffusion, electrochemical transport, and other irreversible processes. The results provide a unified thermodynamic geometric basis for transport equations, clarify the conditions under which continuum models remain valid, and offer a new interpretation of turbulence scaling and phase-transition-induced breakdown.
	
\end{abstract}

\section{Introduction}

The mathematical description of irreversible transport processes has historically developed along two largely independent tracks. 
Equilibrium thermodynamics provides a geometric structure based on convex potentials, Legendre transforms, and Hessian-derived metrics, forming the basis of information and fluctuation geometry \cite{ruppeiner1979thermodynamics,weinhold1975metric,callen1985thermodynamics}. 
Separately, continuum transport theories such as the Navier--Stokes equations, heat conduction, diffusion, and electrochemical transport describe how macroscopic fluxes evolve in physical space \cite{onsager1931reciprocal1,degrootmazur1984}. 
Although both frameworks originate from the same thermodynamic principles, they are rarely formulated within a common mathematical structure: the equilibrium geometry characterizes admissible states, while transport equations govern dynamics.

This separation raises a fundamental question: 
\begin{center}
	\emph{Can transport dynamics be expressed as geometric evolution equations on thermodynamic state space, such that constitutive laws and dissipation emerge from the geometry rather than being imposed externally?}
\end{center}

Progress in this direction has appeared in several contexts---including Onsager's variational formulation of irreversible processes \cite{onsager1931reciprocal2}, Rayleigh's dissipation function, contact and Riemannian thermodynamic geometry, and recent work relating curvature to fluctuations and irreversibility \cite{crooks2007geometry,salasnich2015geometric}. 
However, a framework that simultaneously incorporates (i) the equilibrium state-space geometry and (ii) the local transport structure into a single dynamical description has not yet been constructed.

The present work develops such a unified formulation. 
The key observation is that irreversible systems naturally possess two complementary geometric structures:

\begin{enumerate}
	\item A \textbf{thermodynamic state-space metric}, obtained from the Hessian of a thermodynamic potential. 
	When the enthalpy is expressed in coordinates where entropy is explicit, the resulting metric acquires a Lorentzian-type signature: entropy acts as a time-like coordinate, while its conjugate (e.g., pressure) plays the role of a spatial-like coordinate.
	
	\item A \textbf{local transport metric}, arising from Onsager reciprocity. 
	The inverse Onsager matrix defines a positive-definite quadratic dissipation function and therefore a metric on the space of fluxes or velocity gradients.
\end{enumerate}

A central element linking these two structures is the introduction of \textbf{thermodynamic proper time}. 
Entropy production provides a natural intrinsic clock for irreversible processes, allowing laboratory time $t$ to be reparameterized by a monotonic quantity $\tau$ via:
\begin{equation}
	d\tau^2 = \sigma\, dt^2,
	\label{eq:thermo_time}
\end{equation}
where $\sigma$ is the local entropy production rate. 
This identification connects the evolution on the thermodynamic manifold to physical transport dynamics and provides a common parametrization for both geometries. 
In this formulation, the progression of an irreversible process is measured not only by external time but by accumulated dissipation.

When the state-space metric, the transport metric, and thermodynamic proper time are combined within a single variational principle, the resulting geodesic equations reproduce familiar continuum transport laws. 
For an incompressible Newtonian fluid, the Euler--Lagrange equations reduce to the Navier--Stokes equations without requiring a separately imposed constitutive law for viscosity. 
Furthermore, the balance between inertial curvature and dissipation expressed through the relationship between $t$ and $\tau$ yields the Kolmogorov microscale as a minimal geometric length scale, providing a geometric interpretation of turbulence cascade termination.

The framework also clarifies the origin of breakdown conditions in continuum models. 
Whereas classical treatments interpret singularities as analytic failure of governing equations, the present formulation identifies them as degeneracies of the thermodynamic metric---for example, at cavitation or critical points where compressibility diverges and the underlying manifold changes topology.

Although the full derivation is demonstrated for fluid flow, the formulation is not fluid-specific. 
By selecting an appropriate thermodynamic potential and Onsager matrix, the same geometric structure applies to heat transport, mass diffusion, electrochemical systems, and other irreversible processes.

The purpose of this work is not to replace existing transport equations, but to provide a unifying geometric framework that links thermodynamic state structure, dissipation, and dynamical evolution. 
This approach offers a new perspective on stability, scaling, and breakdown in irreversible systems and suggests a foundation for generalized transport theories applicable across physical domains.

\section{Background and Motivation}

The study of irreversible processes spans statistical physics, information theory, and continuum mechanics. 
Classical equilibrium thermodynamics provides a well-defined geometric structure through Legendre transforms, Maxwell relations, and convexity of thermodynamic potentials \cite{callen1985thermodynamics}. 
In parallel, non-equilibrium thermodynamics and continuum transport theory describe the evolution of macroscopic fields such as velocity, temperature, concentration, or chemical potential \cite{degrootmazur1984}. 
Although both perspectives originate from the same microscopic foundations, the relationship between thermodynamic structure and macroscopic dynamics remains conceptually and mathematically indirect.

\subsection{Thermodynamic Geometry}

Geometric approaches to equilibrium thermodynamics have demonstrated that thermodynamic potentials naturally induce Riemannian metrics on the space of equilibrium states. 
Examples include Weinhold's energy metric \cite{weinhold1975metric}, Ruppeiner's entropy metric \cite{ruppeiner1979thermodynamics}, and formulations based on contact and information geometry \cite{mrugala1991contact, crooks2007geometry}. 
These metrics capture stability properties through definiteness of response functions and reveal curvature singularities at critical points.

However, existing thermodynamic geometries typically lack a built-in notion of dynamics: they describe the landscape of admissible equilibrium states rather than how systems evolve through that space under driving or dissipation.

\subsection{Onsager Reciprocity and Dissipation Structure}

Near equilibrium, irreversible processes obey linear relationships between generalized fluxes $J^i$ and thermodynamic forces $X_i$, as formulated by Onsager \cite{onsager1931reciprocal1, onsager1931reciprocal2}:
\begin{equation}
	J^i = \sum_j L^{ij} X_j,
\end{equation}
where $L^{ij}$ is a symmetric positive-definite transport matrix. 
The corresponding Rayleigh dissipation function defines a quadratic cost for irreversible motion,
\begin{equation}
	\Phi = \frac{1}{2} \sum_{ij} R_{ij} J^i J^j,
	\qquad R = L^{-1}.
\end{equation}
This structure provides a natural metric on the space of transport processes and forms the basis for variational formulations of irreversible dynamics \cite{prigogine1967thermodynamics, mazenko2006non}.

While Onsager theory offers a local linear mapping between forces and fluxes, it does not by itself specify the global geometry of thermodynamic transitions or how transport equations relate to thermodynamic stability metrics.

\subsection{Continuum Transport Models and Their Limitations}

In continuum mechanics, constitutive relations are typically introduced externally to close the governing balance equations. 
For fluids, the incompressible Navier--Stokes equations serve as a standard model for momentum transport, while analogous laws describe heat conduction, diffusion, or electrochemical transport. 
Despite their success, several open questions remain:

\begin{itemize}
	\item Navier--Stokes is not normally derived from thermodynamic first principles; viscosity and stress closure are imposed rather than geometrically generated.
	\item The role of thermodynamic stability (e.g., heat capacity, compressibility) in determining dynamical behavior is indirect.
	\item The emergence of characteristic scales in turbulent flow, such as the Kolmogorov length, is phenomenological rather than structural.
	\item The breakdown of continuum models (e.g., cavitation, phase transitions) is diagnosed empirically rather than as a geometric singularity.
\end{itemize}

These gaps suggest the need for a framework in which thermodynamic structure and irreversible transport dynamics are not introduced separately, but arise from a single geometric formulation.

\subsection{Motivation for a Unified Geometric Framework}

A conceptual bridge between thermodynamic geometry and transport dynamics would offer several advantages:

\begin{enumerate}
	\item It could provide a first-principles derivation of transport equations, reducing reliance on empirical constitutive assumptions.
	\item It may reveal geometric constraints that govern scaling behavior and regularity, particularly in non-linear systems such as turbulence.
	\item It could offer a rigorous way to determine when a continuum model is expected to break down, based on degeneracy of the thermodynamic metric rather than analytic singularity.
	\item It would allow the treatment of fluids, heat conduction, diffusion, and electrochemical systems within a common mathematical structure by selecting appropriate thermodynamic coordinates and Onsager relations.
\end{enumerate}

The present work develops such a structure by combining a thermodynamic state-space metric with a dissipation-based transport metric into a unified action-based geometric formalism. 
The following sections construct this framework explicitly and demonstrate how classical transport laws, including the Navier--Stokes equations, emerge as geodesic flows on this thermodynamic manifold.

\section{Thermodynamic State-Space Geometry}
\label{sec:state_geometry}

The first component of the proposed framework is a geometric description of the thermodynamic state space. 
Following established approaches to thermodynamic geometry \cite{weinhold1975metric,ruppeiner1979thermodynamics}, we construct a metric tensor from the Hessian of a thermodynamic potential expressed in its natural variables. 
However, unlike previous formulations, we explicitly seek a representation in which entropy appears as a coordinate of the manifold so that the resulting metric possesses a time-like structure reflecting the directional nature of entropy production.

\subsection{Choice of Thermodynamic Potential}

To ensure that entropy appears explicitly as a coordinate, we select the enthalpy,
\begin{equation}
	H = U + PV,
\end{equation}
expressed in its natural variables $(S,P)$.
The differential form of the enthalpy reads
\begin{equation}
	dH = T\, dS + V\, dP,
	\label{eq:enthalpy_firstlaw}
\end{equation}
where $T$ is temperature and $V$ is volume. 
This representation plays a central role in systems where mechanical work and pressure fluctuations are relevant, such as fluids and compressible media, although nothing in the construction restricts the framework to such systems.

\subsection{Hessian Metric}

We define the thermodynamic state-space metric as the Hessian of $H$
\begin{equation}
	g_{\mu\nu} = \frac{\partial^2 H}{\partial x^\mu \partial x^\nu},
	\qquad x^\mu \in \{S,P\}.
	\label{eq:state_metric_def}
\end{equation}
From Eq.~\eqref{eq:enthalpy_firstlaw}, the first derivatives are
\begin{equation}
	\frac{\partial H}{\partial S} = T, 
	\qquad 
	\frac{\partial H}{\partial P} = V.
\end{equation}
Differentiating again yields the diagonal metric components
\begin{align}
	g_{SS} &= \left(\frac{\partial T}{\partial S}\right)_P, \\
	g_{PP} &= \left(\frac{\partial V}{\partial P}\right)_S.
\end{align}

These derivatives can be expressed in terms of measurable thermodynamic response functions. 
The heat capacity at constant pressure is defined by
\begin{equation}
	C_P = T \left(\frac{\partial S}{\partial T}\right)_P
	\quad\Rightarrow\quad
	\left(\frac{\partial T}{\partial S}\right)_P = \frac{T}{C_P},
\end{equation}
and the adiabatic compressibility $\kappa_S$ satisfies
\begin{equation}
	\kappa_S
	= -\frac{1}{V}\left(\frac{\partial V}{\partial P}\right)_S
	\quad\Rightarrow\quad
	\left(\frac{\partial V}{\partial P}\right)_S = -V\,\kappa_S.
\end{equation}
Thus the metric takes the form
\begin{equation}
	g_{SS} = \frac{T}{C_P},
	\qquad
	g_{PP} = - V \kappa_S.
	\label{eq:metric_components}
\end{equation}

\subsection{Signature and Interpretation}

For all thermodynamically stable single-phase systems, $T>0$, $C_P>0$, $\kappa_S>0$, and $V>0$. 
Therefore,
\begin{equation}
	g_{SS}>0,
	\qquad
	g_{PP}<0,
\end{equation}
and the metric has Lorentzian-type signature $(+,-)$ (up to a conventional overall sign). 
The resulting line element is
\begin{equation}
	ds^2 
	= \left(\frac{T}{C_P}\right)dS^2 
	- (V\kappa_S)\, dP^2.
	\label{eq:state_line_element}
\end{equation}

The positive component associated with entropy identifies $S$ as a \emph{time-like coordinate}: irreversible processes evolve in the direction of increasing entropy, and the second law ensures monotonicity. 
The negative component associated with pressure identifies $P$ as a \emph{spatial-like coordinate} measuring mechanical deformation or stress in the thermodynamic state space.

\subsection{Thermodynamic Proper Time}
\label{sec:proper_time_state}

The identification of entropy as the time-like coordinate in the state-space metric suggests that thermodynamic processes may be parametrized by an intrinsic notion of time. 
Since the second law guarantees that entropy increases monotonically along any irreversible trajectory, entropy provides a natural ordering parameter. 
To make this correspondence precise, we introduce a proper-time parametrization associated with displacements along the entropy direction of the metric.

Restricting the line element~\eqref{eq:state_line_element} to variations in entropy at fixed pressure yields
\begin{equation}
	ds_{\parallel S}^2 
	= \left(\frac{T}{C_P}\right) dS^2 .
	\label{eq:entropy_line}
\end{equation}
Since $g_{SS}>0$, the square root of this expression defines a natural time-like parameter,
\begin{equation}
	d\tau = \sqrt{\frac{T}{C_P}}\, dS .
	\label{eq:tau_def_state}
\end{equation}
We refer to $\tau$ as the \emph{thermodynamic proper time}: it measures progression along the time-like direction of the thermodynamic manifold.  

Equation~\eqref{eq:tau_def_state} shows that the rate at which thermodynamic time accumulates depends not only on entropy change but also on the local thermodynamic response encoded in the metric coefficients. 
A given change in entropy contributes more to the elapsed thermodynamic time when $C_P$ is small or temperature is large; conversely, near critical points where $C_P\to\infty$, the increment $d\tau$ becomes vanishingly small, reflecting the slowing down of equilibration characteristic of critical phenomena.

Although $\tau$ is introduced here purely as a geometric parameter, it will acquire dynamical significance once transport processes are included. 
In particular, when entropy becomes a dynamical field $S(x,t)$ evolving in laboratory time $t$, the relation~\eqref{eq:tau_def_state} connects the state-space geometry to entropy production rates and provides a parametrization common to both thermodynamic and transport geometries. 
This role makes $\tau$ a key structural element in unifying the equilibrium metric and the dissipation-induced local transport geometry developed in the next section.

\subsection{Null Curves and Characteristic Propagation}

Null curves satisfy $ds^2 = 0$, yielding
\begin{equation}
	\left(\frac{dP}{dS}\right)^2 
	= \frac{T}{V\,C_P\,\kappa_S}.
	\label{eq:null_speed}
\end{equation}
We define the thermodynamic characteristic speed
\begin{equation}
	c_{\mathrm{th}}^2 = \frac{T}{V\,C_P\,\kappa_S}.
\end{equation}
For simple equations of state, this quantity is proportional to the adiabatic sound speed, implying that the causal structure of the thermodynamic state space is tied to the propagation of pressure disturbances in the material.

\begin{figure}[btbp]
	\centering
	\includegraphics[width=0.8\textwidth]{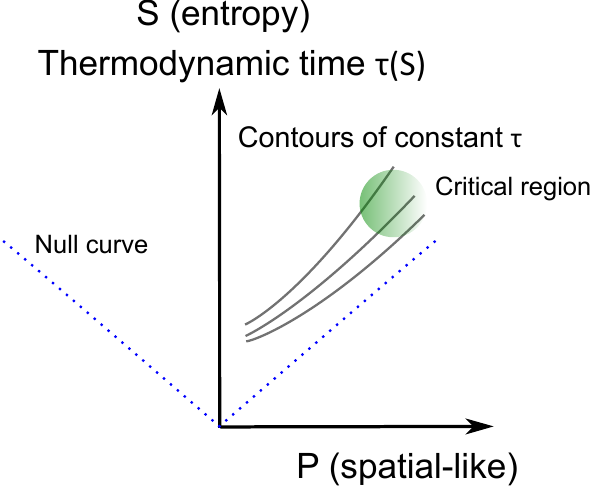}
	
	\caption{Schematic representation of the thermodynamic manifold. 
		Entropy defines a time-like direction with proper time $d\tau=\sqrt{T/C_P}\,dS$, 
		while pressure spans a spatial-like axis. 
		Null curves $ds^2=0$ define thermodynamic causality via $c_{\mathrm{th}}^2=T/(VC_P\kappa_S)$. 
		Contours of constant $\tau$ spread apart (become sparse) near critical points where $C_P\to\infty$, 
		reflecting the divergence of the metric and the physical slowing of relaxation.}
	
	\label{fig:fig1_thermo_manifold}
\end{figure}

Equation~\eqref{eq:null_speed} therefore establishes a causal interpretation: the geometry permits only trajectories whose entropy-pressure motion respects the characteristic bound set by $c_{\mathrm{th}}$, analogous to light-cone structure in Lorentzian geometry (see figure \ref{fig:fig1_thermo_manifold}). 
We emphasize that this correspondence is conceptual rather than literal; the analogy is introduced as an interpretive tool rather than a statement of physical equivalence to spacetime geometry.

 \bigskip

The thermodynamic state metric~\eqref{eq:state_line_element} provides the global geometry upon which transport dynamics will later be defined. 
In the next section, we introduce the second geometric structure: a dissipation-based metric derived from Onsager reciprocity, which governs local irreversible evolution.

\section{Local Transport Geometry from Linear Irreversible Thermodynamics}
\label{sec:transport_geometry}

The thermodynamic state-space metric introduced in Sec.~\ref{sec:state_geometry} characterizes stability and admissible directions of change, but it does not specify how a system evolves along these directions. 
To describe irreversible dynamics, we introduce a complementary geometric structure based on linear irreversible thermodynamics. 
In this formulation, fluxes and forces are connected through Onsager reciprocity, and the dissipation associated with irreversible motion induces a metric on the space of transport processes.

\subsection{Fluxes, Forces, and Onsager Reciprocity}

Near equilibrium, irreversible processes satisfy the linear relation \cite{onsager1931reciprocal1,onsager1931reciprocal2}
\begin{equation}
	J^i = \sum_j L^{ij} X_j,
	\label{eq:Onsager_linear_relation}
\end{equation}
where the generalized forces $X_j$ are gradients of intensive variables and $L^{ij}$ is a symmetric positive-definite Onsager matrix.

While Eq.~\eqref{eq:Onsager_linear_relation} defines kinematics, the energetic cost of irreversible evolution is encoded in entropy production.

\subsection{Rayleigh Dissipation Function and Entropy Production}

The irreversible entropy production rate is the bilinear form
\begin{equation}
	\sigma = \sum_i J^i X_i \ge 0 ,
	\label{eq:sigma_basic}
\end{equation}
consistent with the second law. 
Using Eq.~\eqref{eq:Onsager_linear_relation}, this may be written as the quadratic Rayleigh dissipation function
\begin{equation}
	\sigma = \sum_{ij} R_{ij} J^i J^j ,
	\qquad R = L^{-1},
	\label{eq:entropy_quad}
\end{equation}
where $R_{ij}$ is the resistivity tensor. 
Equation~\eqref{eq:entropy_quad} provides a natural inner product on flux space and defines the \textit{transport metric}
\begin{equation}
	ds_{\mathrm{diss}}^2 = R_{ij} J^i J^j \, dt^2 .
	\label{eq:transport_metric}
\end{equation}

\subsection{Thermodynamic Proper Time and Dissipation}
\label{sec:tau_transport}

The transport metric defined in Eq.~\eqref{eq:transport_metric} links irreversible fluxes to the progression of time along a thermodynamic trajectory.
The irreversible entropy production rate is defined by
\begin{equation}
	\sigma = \frac{dS}{dt} = R_{ij} J^i J^j \ge 0,
	\label{eq:sigma_def}
\end{equation}
consistent with the Second Law.
Motivated by the role of $\sigma$ as the driver of irreversible evolution, we introduce a \emph{dynamical thermodynamic time} parametrization based on dissipation:
\begin{equation}
	d\tau_{\mathrm{phys}}^2 = \sigma\, dt^2,
	\qquad \text{or equivalently} \qquad
	d\tau_{\mathrm{phys}} = \sqrt{\sigma}\, dt .
	\label{eq:tau_def}
\end{equation}
This definition measures the accumulated irreversible change along a process.
Unlike laboratory time $t$, which advances uniformly, $\tau_{\mathrm{phys}}$ advances only when dissipation occurs and vanishes at equilibrium.

To understand the geometric implication of this parameter, we recall the \emph{geometric proper time} defined in Sec.~\ref{sec:state_geometry} (Eq.~\eqref{eq:tau_def_state}), which measures distance on the equilibrium manifold:
\begin{equation}
	d\tau_{\mathrm{geom}} = \sqrt{\frac{T}{C_P}}\, dS .
\end{equation}
While $d\tau_{\mathrm{geom}}$ depends solely on equilibrium response functions (where the system \emph{is}), $d\tau_{\mathrm{phys}}$ depends on the local dissipation rate (how fast the system \emph{moves}).

Combining these definitions using $dS = \sigma\, dt$ reveals the coupling between the static geometry and dynamic transport:
\begin{equation}
	d\tau_{\mathrm{phys}} = \sqrt{\frac{\sigma C_P}{T}}\, d\tau_{\mathrm{geom}}.
	\label{eq:time_relation}
\end{equation}
The proportionality term serves as a \emph{thermodynamic time dilation factor}:
\begin{equation}
	\Lambda = \sqrt{\frac{\sigma C_P}{T}} = \sqrt{\frac{C_P}{T} R_{ij}J^i J^j}.
\end{equation}
Equation~\eqref{eq:time_relation} shows that the two parametrizations coincide only when $\Lambda = 1$.
This distinction offers a physical interpretation of the flow regimes:
\begin{itemize}
	\item \textbf{Near-equilibrium ($\Lambda \ll 1$):} Dissipation is weak. One unit of geometric progression ($d\tau_{\mathrm{geom}}$) requires substantial laboratory time. The system explores the state space slowly.
	\item \textbf{Highly dissipative ($\Lambda \gg 1$):} Dissipation is strong. The dynamical time advances rapidly relative to the geometric distance, effectively ``compressing'' the thermodynamic evolution.
\end{itemize}

As we will demonstrate in Sec.~\ref{sec:kolmogorov}, the condition $\Lambda \sim 1$ (where the geometric and dynamical time scales balance) corresponds to the Kolmogorov microscale in turbulence.
Thus, $\Lambda$ acts as the bridge between the equilibrium metric structure and the transport dynamics derived in the following section.

\paragraph{Remark on Onsager Linearity.}
Although the transport relations employed here are linear in the Onsager sense, this linearity should not be interpreted in the traditional restriction to weakly driven or near-equilibrium processes in laboratory time. In the present framework, Onsager reciprocity is tied to the thermodynamic state manifold through the introduction of thermodynamic proper time. As a result, irreversible evolution remains locally linear along the thermodynamic manifold even when the system is strongly driven or far from equilibrium when viewed in laboratory time. Large gradients and intense dissipation manifest not as nonlinear transport laws, but through the reparametrization of time itself via entropy production. The relevant limitation is therefore not the magnitude of driving, but the validity of local thermodynamic equilibrium and quadratic entropy production, assumptions that remain appropriate throughout single-phase continuum dynamics.

\subsection{Structure of the Transport Metric}

The resistivity tensor $R_{ij}$ may be scalar, diagonal, anisotropic, or coupled depending on the physical system. Examples include:
\begin{itemize}
	\item thermal conduction: $R = 1/\kappa_T$,
	\item isotropic Newtonian viscosity: $R \sim \mu$,
	\item coupled thermo-electric or electrochemical transport.
\end{itemize}

As long as $R_{ij}$ remains positive-definite, the transport metric is well-defined and the system evolves monotonically in $\tau$. 
Loss of definiteness corresponds to breakdown of the continuum description and will be examined in Sec.~\ref{sec:breakdown}.

\subsection{Toward a Dynamical Principle}

The unified geometric structure is now complete: the thermodynamic state metric describes admissible directions and curvature of the equilibrium manifold, while the transport metric determines the rate at which the system advances through that manifold in thermodynamic time. 
In the next section, we construct a variational principle that combines these two structures and show that its Euler--Lagrange equations reproduce standard continuum transport laws.

\section{Geometric Derivation of Transport Dynamics}
\label{sec:dynamics}

With the thermodynamic state-space geometry and transport metric established, we now formulate irreversible dynamics as a variational problem on the combined manifold. 
The goal is to determine whether macroscopic transport equations arise as geodesic evolution equations when the system is parameterized by thermodynamic proper time rather than externally imposed laboratory time.

\subsection{Thermodynamic Action Functional}

Motivated by Onsager’s variational formulation and by the structure of analytical mechanics, we introduce a thermodynamic action functional. 
However, unlike conventional formulations, the action is written using the thermodynamic time parameter $\tau$ introduced in Eq.~\eqref{eq:tau_def}. 

We define
\begin{equation}
	\mathcal{A} = \int_{\tau_1}^{\tau_2}
	\left[
	\mathcal{T}
	- \mathcal{U}
	- \Phi
	\right]\, d\tau ,
	\label{eq:thermo_action_tau}
\end{equation}
where:
\begin{itemize}
	\item $\mathcal{T}$ represents inertial kinetic energy,
	\item $\mathcal{U}$ encodes thermodynamic constraints arising from the state metric,
	\item $\Phi$ corresponds to dissipation weighted by the transport metric.
\end{itemize}

The change of parameter from $t$ to $\tau$ introduces the thermodynamic time-scaling factor:
\begin{equation}
	\frac{dt}{d\tau} 
	= \left(
	\frac{C_P}{T \, \sigma}
	\right)^{1/2},
	\qquad
	\sigma = R_{ij}J^iJ^j .
	\label{eq:reparam_relation}
\end{equation}
Equation~\eqref{eq:reparam_relation} couples dissipation directly to the rate at which the system traverses the thermodynamic manifold.

\subsection{Application to Continuum Fluids}

For a Newtonian, incompressible fluid with constant density $\rho$, the generalized coordinate is the velocity field $\mathbf{u}(\mathbf{x},\tau)$.

The kinetic energy density expressed in thermodynamic time becomes:
\begin{equation}
	\mathcal{T} = \frac{1}{2} \rho\, 
	\left(\frac{d\mathbf{u}}{d\tau}\right)^2
	\left(\frac{dt}{d\tau}\right)^2
	= \frac{1}{2}\rho \mathbf{u}^2 
	\left(\frac{dt}{d\tau}\right)^2.
\end{equation}

The incompressibility constraint appears via a Lagrange multiplier:
\begin{equation}
	\mathcal{U} = p\,(\nabla\!\cdot\!\mathbf{u}).
\end{equation}

Dissipation enters through the transport metric as
\begin{equation}
	\Phi =
	\frac{1}{2}\mu
	\left( \nabla\mathbf{u} + \nabla\mathbf{u}^{T} \right)^2
	\left(\frac{dt}{d\tau}\right). \footnotemark
\end{equation}
\footnotetext{
	The factor $\left(dt/d\tau\right)$ multiplying the dissipation term arises from reparametrizing the Onsager–Rayleigh dissipation functional, which is originally defined per unit \emph{laboratory time}. During the change of variables from $t$ to thermodynamic proper time $\tau$, the action measure transforms as $dt = (dt/d\tau)\, d\tau$. Unlike the kinetic term, the dissipation functional contains no explicit time derivatives and therefore acquires only a single power of this factor. This distinction encodes the physical asymmetry between reversible transport (quadratic in time parametrization) and irreversible entropy production (defined per physical time) and reflects the fact that dissipation governs the rate at which thermodynamic time advances relative to laboratory time.
}

Using Eq.~\eqref{eq:reparam_relation}, the full Lagrangian density becomes
\begin{equation}
	\mathcal{L}
	= \frac{1}{2}\rho \mathbf{u}^2 
	\left(\frac{dt}{d\tau}\right)^2
	- p(\nabla\!\cdot\!\mathbf{u})
	- \frac{1}{2}\mu
	\left(\nabla\mathbf{u} + \nabla\mathbf{u}^T\right)^2
	\left(\frac{dt}{d\tau}\right).
	\label{eq:full_lagrangian_tau}
\end{equation}

Thus, the dissipation term does not merely penalize motion—it modifies the effective parametrization of time itself.

\subsection{Variational Derivation}

The Euler--Lagrange equation in thermodynamic time is
\begin{equation}
	\frac{d}{d\tau}
	\left(
	\frac{\partial \mathcal{L}}
	{\partial \left( d\mathbf{u}/d\tau \right)}
	\right)
	+
	\nabla\cdot\left(
	\frac{\partial \mathcal{L}}{\partial (\nabla\mathbf{u})}
	\right)
	-
	\frac{\partial \mathcal{L}}{\partial \mathbf{u}}
	= 0 .
	\label{eq:EL_tau}
\end{equation}

Converting derivatives using Eq.~\eqref{eq:reparam_relation} yields, after simplification,
\begin{equation}
	\rho\left(
	\frac{\partial \mathbf{u}}{\partial t}
	+ \mathbf{u}\cdot\nabla\mathbf{u}
	\right)
	= -\nabla p + \mu\nabla^2\mathbf{u},
\end{equation}
with the constraint
\[
\nabla\cdot\mathbf{u}=0.
\]

Thus, the Navier--Stokes equations emerge from a geodesic principle parameterized by thermodynamic proper time.

\subsection{Interpretation}

This formulation provides a new physical interpretation:

\begin{quote}
	\textit{Inertia governs how momentum evolves in laboratory time $t$, while dissipation governs how entropy advances thermodynamic time $\tau$. Navier--Stokes emerges at the intersection of these two temporal structures.}
\end{quote}

Regions of strong gradients (large $\sigma$) evolve rapidly in $\tau$, while near-equilibrium regions evolve slowly. 
Later, in Sec.~\ref{sec:kolmogorov}, we show that the condition $d\tau/dt \sim 1$ yields the Kolmogorov microscale as a geometric crossover from inertial to dissipative behavior.

\section{Turbulent Scaling and the Kolmogorov Length as a Geometric Cutoff}
\label{sec:kolmogorov}

The geometric formulation developed in Secs.~\ref{sec:state_geometry}--\ref{sec:dynamics} establishes that two competing mechanisms govern the evolution of fluid motion: inertial curvature, which drives deformation in laboratory time, and dissipation, which governs the rate at which thermodynamic proper time progresses.
Turbulence expresses the competition between these two temporal structures across scales.
In this section, we show that the Kolmogorov length scale emerges as the point where the two time parameters become commensurate, leading to a natural lower bound on fluid motion.

\subsection{Scaling of Inertial Contributions}

Consider eddies of characteristic size $\ell$ with velocity scale $u_\ell$.
Kolmogorov's phenomenological scaling asserts that the rate of cascade of kinetic energy $\epsilon$ is approximately constant across scales \cite{kolmogorov1941local}:
\begin{equation}
	\epsilon \sim \frac{u_\ell^3}{\ell}.
\end{equation}
Solving for the characteristic velocity yields $u_\ell \sim (\epsilon \ell)^{1/3}$.
Under these assumptions, the characteristic inertial acceleration associated with curvature of the dynamical manifold scales as
\begin{equation}
	a_{\mathrm{inertial}}(\ell)
	\sim \frac{u_\ell^2}{\ell}
	\sim \epsilon^{2/3} \ell^{-1/3}.
	\label{eq:inertial_scaling}
\end{equation}
This contribution grows with decreasing $\ell$, reflecting increasingly rapid deformation at smaller turbulent scales.

\subsection{Scaling of Dissipation and Entropy-Time Weighting}

The dissipation metric introduced in Sec.~\ref{sec:transport_geometry} assigns energetic cost to velocity gradients.
For a Newtonian fluid, the local dissipation rate scales as
\begin{equation}
	\sigma(\ell) 
	\sim \mu \left(\frac{u_\ell}{\ell}\right)^2
	\sim \mu \epsilon^{2/3} \ell^{-4/3}.
\end{equation}
Through the thermodynamic time relation established in Eq.~\eqref{eq:reparam_relation}, dissipation modifies the parametrization of evolution:
\begin{equation}
	\frac{d\tau}{dt} 
	\sim \sqrt{\sigma}
	\sim \sqrt{\mu}\,\epsilon^{1/3} \ell^{-2/3}.
	\label{eq:tau_vs_t_scale}
\end{equation}
Equation~\eqref{eq:tau_vs_t_scale} expresses a key qualitative fact: \textit{small scales evolve faster in thermodynamic time than in laboratory time}.
In the inertial range ($\ell \gg \eta$), dissipation is weak ($d\tau/dt \ll 1$) and dynamics appear nearly reversible.
At sufficiently small scales, dissipation dominates ($d\tau/dt \gg 1$) and motion is rapidly converted into entropy.

\subsection{Balance Condition and the Kolmogorov Length}

The Kolmogorov scale corresponds to the point where inertial curvature and dissipation-induced time dilation become comparable.
Equivalently, it is where the two time parameters advance at the same rate:
\begin{equation}
	\frac{d\tau}{dt} \sim 1.
	\label{eq:balance_condition}
\end{equation}
This geometric condition is physically equivalent to the balance between the inertial curvature generation and the viscous transport stiffness.
From the Navier-Stokes terms, the viscous restoring acceleration scales as $a_{\mathrm{viscous}} \sim \nu u_\ell / \ell^2$.
Equating the inertial and viscous contributions yields:
\begin{equation}
	\underbrace{\frac{u_\ell^2}{\ell}}_{\text{Inertia}} 
	\sim 
	\underbrace{\nu \frac{u_\ell}{\ell^2}}_{\text{Dissipation}}
	\quad\Rightarrow\quad
	\frac{u_\ell \ell}{\nu} \sim 1.
\end{equation}
Substituting the time-scaling Eq.~\eqref{eq:tau_vs_t_scale} into the balance condition Eq.~\eqref{eq:balance_condition} recovers the standard Kolmogorov length:
\begin{equation}
	1 \sim \sqrt{\mu}\,\epsilon^{1/3} \ell^{-2/3}
	\quad\Rightarrow\quad
	\boxed{
		\eta = \left(\frac{\nu^3}{\epsilon}\right)^{1/4}.
	}
\end{equation}
In this framework, $\eta$ is not an empirical cutoff but the geometric fixed point where inertial curvature and dissipation-induced thermodynamic timing balance.

\subsection{Interpretation as a Minimal Geometric Resolution}

For $\ell > \eta$, the factor $d\tau/dt \ll 1$ and the dominant contribution to the action comes from inertia.
Fluid motion proceeds along nearly reversible geodesics, and eddies behave as dynamically coherent structures.
For $\ell < \eta$, the dissipation metric dominates and $d\tau/dt \gg 1$.
Further creation of curvature becomes exponentially costly in thermodynamic time, and motion is suppressed: the system transitions from dynamical evolution to irreversible entropy production.
Thus, the Kolmogorov length marks the boundary between two geometric regimes:
\begin{equation}
	\begin{split}
		\ell > \eta \;\Rightarrow\; \text{inertial, curvature-generating motion},\\
		\ell < \eta \;\Rightarrow\; \text{viscous, entropy-dominated smoothing}.
	\end{split}
\end{equation}

\subsection{Reynolds Number as a Temporal Stability Ratio}
\label{sec:reynolds_transition}

The analysis in the preceding sections shows that the transition from reversible inertial motion to dissipative smoothing depends on the relative rate at which the system evolves in laboratory time $t$ versus thermodynamic time $\tau$. 
At large scales, dissipation is weak and $d\tau/dt \ll 1$; flow evolves predominantly under inertia. 
At sufficiently small scales, dissipation dominates and $d\tau/dt \gg 1$. 
The Reynolds number characterizes this competition at the system scale.

To formalize this, consider two neighboring fluid trajectories separated by $\xi^\mu$. 
Their stability is governed by the geodesic deviation equation
\begin{equation}
	\frac{D^2 \xi^\mu}{d\tau^2} 
	= -R^\mu_{\ \nu\alpha\beta} U^\nu \xi^\alpha U^\beta,
	\label{eq:jacobi}
\end{equation}
where $R^\mu_{\ \nu\alpha\beta}$ is the curvature associated with the transport metric.  

The curvature term scales as
\begin{equation}
	R \sim \frac{1}{\nu} (\nabla u)^2,
\end{equation}
reflecting the fact that curvature increases when viscosity is small and gradients are large.

Using characteristic velocity $u$ and length scale $L$, the inertial contribution to trajectory divergence scales as
\begin{equation}
	a_{\mathrm{inertial}} \sim \frac{u^2}{L},
\end{equation}
while the dissipation-induced restoring term associated with transport metric stiffness scales as
\begin{equation}
	a_{\mathrm{diss}} \sim \nu \frac{u}{L^2}.
\end{equation}

The boundary between coherent (laminar) and unstable (turbulent) dynamics occurs at neutral stability,
\begin{equation}
	a_{\mathrm{inertial}} \approx a_{\mathrm{diss}}.
\end{equation}
Substituting the scalings yields
\begin{equation}
	\frac{u^2/L}{\nu u/L^2}
	= \frac{uL}{\nu}
	= \mathrm{Re},
\end{equation}
identifying the Reynolds number as the dimensionless ratio governing the sign and magnitude of dynamical curvature:
\begin{equation}
	\mathrm{Re} \gtrsim \mathrm{Re}_{\mathrm{crit}}
	\quad \Rightarrow \quad
	\frac{d\tau}{dt} \text{ increases and geodesics diverge exponentially.}
\end{equation}

In this formulation, the Reynolds number admits a natural geometric interpretation:

\begin{equation}
	\boxed{
		\mathrm{Re} 
		= \frac{\text{inertial time scale}}{\text{dissipative entropy time scale}}
		= \frac{t_{\text{inertial}}}{t_{\text{irreversible}}}.
	}
\end{equation}

Thus, Reynolds number serves as the \emph{macroscopic analogue} of the Kolmogorov balance condition $d\tau/dt \sim 1$: turbulence emerges when inertial dynamics advance more rapidly in laboratory time than dissipation can advance in thermodynamic time. 
Laminar flow corresponds to the opposite regime, where entropy production dominates and suppresses curvature growth in the transport manifold.

\subsection{Summary}

In this interpretation, the Kolmogorov scale arises naturally from the coupling between the state-space geometry and the dissipation-induced transport metric. 
It represents the smallest scale at which a meaningful geodesic evolution can occur before the manifold forces a transition from reversible motion in laboratory time to irreversible dissipation in thermodynamic time. 
The cutoff is therefore a geometric invariant determined by material properties and energy flux, rather than an externally imposed modeling assumption.

In the next section, we examine how this same geometric structure predicts the breakdown of continuum transport laws when the thermodynamic metric becomes degenerate, such as near cavitation or phase transitions.

\section{Breakdown Conditions and Phase Transitions}
\label{sec:breakdown}

The previous sections demonstrated that transport dynamics arise as geodesic evolution on a manifold equipped with a thermodynamic state metric and a dissipation-based transport metric. 
Within a single thermodynamic phase, this structure enforces smoothness of solutions and introduces a natural minimum scale through the balance between inertial curvature and dissipation. 

However, continuum descriptions are known to fail in certain regimes, including cavitation, critical points, or abrupt structural transitions. 
In this section, we show that such breakdowns correspond to degeneracies of the thermodynamic state metric rather than intrinsic failure of the transport dynamics. 
This provides a geometric criterion for the domain of validity of continuum transport models.

\subsection{Metric Degeneracy and Instability Criteria}

From Eq.~\eqref{eq:metric_components}, the thermodynamic state metric components are
\begin{equation}
	g_{SS} = \frac{T}{C_P}, 
	\qquad 
	g_{PP} = - V\kappa_S.
\end{equation}
Both $C_P$ and $\kappa_S$ serve as thermodynamic stability coefficients: 
$C_P>0$ prevents runaway thermal response, and $\kappa_S>0$ ensures mechanical rigidity against compressive or tensile perturbations.

A breakdown occurs when either response coefficient diverges:
\begin{equation}
	C_P \to \infty 
	\quad\text{or}\quad 
	\kappa_S \to \infty,
\end{equation}
leading to
\begin{equation}
	g_{SS} \to 0,
	\qquad
	g_{PP} \to -\infty.
\end{equation}

At such points, the state-space geometry becomes degenerate: motion in the pressure direction becomes infinitely ``long'' in the metric sense, while the entropy direction loses geometric weight. In differential-geometric terms, the thermodynamic manifold ceases to be a smooth pseudo-Riemannian space.

\subsection{Example: Cavitation}

Cavitation provides a physically transparent example. 
As the liquid pressure approaches the vapor pressure, the system approaches the spinodal boundary, where
\begin{equation}
	\left(\frac{\partial P}{\partial V}\right)_S \to 0,
\end{equation}
and hence
\begin{equation}
	\kappa_S = -\frac{1}{V}\left(\frac{\partial V}{\partial P}\right)_S \to \infty.
\end{equation}

Substitution into the metric gives
\begin{equation}
	g_{PP} \to -\infty,
\end{equation}
indicating that the thermodynamic manifold becomes singular in the pressure direction.

The corresponding acoustic propagation speed,
\begin{equation}
	c_s = \sqrt{\frac{1}{\rho\kappa_S}},
\end{equation}
vanishes as $\kappa_S \to \infty$, implying that the continuum pressure field can no longer propagate causal information. 
At this point, the momentum balance loses physical validity, and the system transitions to a distinct branch of the thermodynamic manifold, corresponding to vapor nucleation.

\begin{figure}[btbp]
	\centering
	\includegraphics[width=0.8\textwidth]{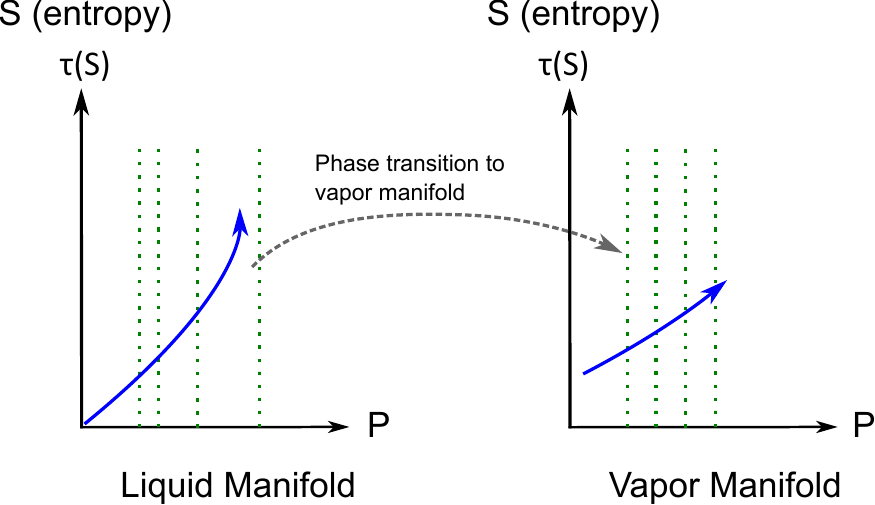}
	
\caption{Schematic of geometric breakdown at cavitation.
	\textbf{(Left) Liquid Manifold:} As the fluid approaches the spinodal limit, the thermodynamic geometry stretches—represented by the \emph{sparse spacing} of the vertical pressure grid lines—due to diverging adiabatic compressibility ($\kappa_S \to \infty$) and the singularity of the metric component $g_{PP} \to -\infty$.
	The fluid trajectory (blue curve) is forced to become vertical as the speed of sound vanishes ($c_s \to 0$), creating an infinite barrier in thermodynamic proper time ($d\tau \to \infty$).
	\textbf{(Center)} The system undergoes a topological transition (gray dashed arrow) across the singular region where the continuum description fails.
	\textbf{(Right) Vapor Manifold:} The trajectory resumes on a distinct, stable manifold with non-degenerate geometry.}
\label{fig:cavitation_breakdown}
	
	\label{fig:fig2_cavitation_jump}
\end{figure}

\subsection{Divergence of Thermodynamic Proper Time}

The irreversible entropy production rate satisfies
\begin{equation}
	\sigma \propto \mu (\nabla \mathbf{u})^2,
\end{equation}
and the coupling between laboratory time $t$ and thermodynamic proper time $\tau$ is
\begin{equation}
	d\tau^2 = \sigma\,dt^2.
\end{equation}

If the system approaches a metric singularity, gradients necessarily diverge:
\begin{equation}
	\nabla \mathbf{u} \to \infty 
	\quad\Rightarrow\quad 
	d\tau \to \infty.
\end{equation}

Thus, although the classical PDE formulation may predict finite-time blow-up in $t$, the thermodynamic proper time required to reach such a state diverges. 
Dynamically, the system asymptotically approaches—yet never reaches—the degenerate manifold. 
The observed ``singularity'' signals a change in phase rather than a breakdown of dynamics.

\subsection{Interpretation of Breakdown}

The results may be summarized as follows:

\begin{enumerate}
	\item \textbf{Within a single, stable thermodynamic phase:}  
	The state metric is non-degenerate and the dissipation metric is positive-definite. 
	Transport takes the form of smooth geodesic evolution.
	
	\item \textbf{At the Kolmogorov scale:}  
	The balance $d\tau/dt \sim 1$ establishes a lower geometric resolution without singularity. 
	Below this scale, motion converts directly into entropy.
	
	\item \textbf{At a phase transition (e.g., cavitation, critical point):}  
	The thermodynamic metric becomes degenerate, the manifold undergoes a change in topology, and the continuum description ceases to apply—not because the governing equations fail mathematically, but because the geometric structure that defines them no longer exists.
\end{enumerate}

\subsection{Summary}

In this framework, breakdown of continuum dynamics is not a pathological feature of the governing equations but a geometric signal that the thermodynamic manifold has changed structure. 
The divergence of the state metric reflects the physical transition to a different phase space, while the divergence of thermodynamic proper time ensures that no finite-time dynamical singularity is physically realizable within the original phase. 
This interpretation provides a natural criterion for the applicability and limits of continuum transport theories.

\section{Generality Beyond Fluids}
\label{sec:generality}

Although the explicit construction in the preceding sections focused on incompressible Newtonian fluids, nothing in the formulation is intrinsically fluid-specific. 
The geometric framework requires only two structural elements:  
(i) a thermodynamic potential defining a state-space metric via its Hessian, and  
(ii) an Onsager matrix defining a dissipation metric on flux space.  
Any irreversible transport process possessing these ingredients admits an equivalent geometric description.

\subsection{Alternative Thermodynamic Potentials}

Different classes of systems employ different natural variables and thermodynamic potentials. Examples include:

\begin{itemize}
	\item Heat conduction and thermal relaxation using $U(S,V)$ or $F(T,V)$,
	\item Chemical diffusion via $G(T,P,\{n_i\})$,
	\item Electrochemical or charge-transport processes involving additional Legendre variables such as electric potential or species chemical potentials.
\end{itemize}

In all cases, the metric is obtained as
\begin{equation}
	g_{\mu\nu} = \frac{\partial^2 \Phi}{\partial x^\mu \partial x^\nu},
\end{equation}
where $\Phi$ is the chosen thermodynamic potential.  
The signature reflects stability coefficients such as heat capacity, compressibility, magnetic susceptibility, or chemical response.

\subsection{Generalized Flux–Force Relations}

Irreversible evolution is described by Onsager-type flux–force relations:
\begin{equation}
	J^i = \sum_j L^{ij}X_j,
\end{equation}
where forces are gradients of intensive variables and fluxes represent generalized transport. Examples include:

\begin{center}
	\begin{tabular}{l|l|l}
		\textbf{Domain} & \textbf{Flux $J^i$} & \textbf{Force $X_i$} \\ \hline
		Heat conduction & $q$ & $\nabla(1/T)$ \\
		Mass diffusion & $J_\alpha$ & $-\nabla \mu_\alpha/T$ \\
		Electrochemistry & $j$ & $E$ \\
		Thermoelectricity & $(q, j)$ & $(\nabla (1/T), E)$ \\
		Viscous flow & Velocity gradients & Stress gradients
	\end{tabular}
\end{center}

The inverse Onsager matrix defines the dissipation metric:
\begin{equation}
	ds_{\mathrm{diss}}^2 = R_{ij}J^iJ^j\,dt^2.
\end{equation}

\subsection{Unified Evolution Principle}

Once both metrics are specified, the evolution law follows from the same variational action:
\begin{equation}
	\delta \int \left( \mathcal{T} - \mathcal{U} - \Phi \right)\, d\tau = 0,
\end{equation}
where $d\tau$ is the thermodynamic proper time defined by the coupling between entropy production and laboratory time.

For fluids this yields the Navier–Stokes equations; for other systems the result may be Fourier’s law, Fickian diffusion, drift–diffusion equations, heat–mass–charge coupled transport, or mixed irreversible evolution equations.

\subsection{Entropy Time as a Universal Dynamical Parameter}

A central feature of the framework is that the thermodynamic proper time,
\begin{equation}
	d\tau = \sqrt{\frac{T}{C_P}}\, dS,
\end{equation}
extends naturally to any irreversible process.  

Once transport processes generate entropy at rate
\begin{equation}
	\sigma = R_{ij}J^iJ^j \ge 0,
\end{equation}
the relation between internal and external time scales becomes
\begin{equation}
	\frac{d\tau}{dt} = \sqrt{\sigma}.
\end{equation}

Thus, entropy time provides a universal clock for irreversible dynamics:  
systems evolve rapidly in $\tau$ where dissipation is large and slowly where gradients are weak or transport vanishes.

Characteristic length or time scales in other domains (thermal diffusion length, Debye screening length, reaction–diffusion scales) emerge as the point where
\begin{equation}
	\frac{d\tau}{dt} \sim 1,
\end{equation}
analogous to the Kolmogorov condition for fluids.

\subsection{Classification of Transport Manifolds}

The combined geometry suggests a natural, domain-independent classification:

\begin{enumerate}
	\item \textbf{Single-phase, finite response:}  
	State metric non-degenerate; smooth geodesic dynamics.
	
	\item \textbf{Multiscale or strongly nonlinear regimes:}  
	Competition between irreversibility and inertia establishes emergent characteristic scales (Kolmogorov, Debye, thermal diffusion, etc.).
	
	\item \textbf{Phase transitions or critical phenomena:}  
	Metric degeneracy signals topological change; evolution proceeds on a different manifold rather than diverging.
\end{enumerate}

\subsection{Summary}

The geometric transport framework developed here is broadly applicable across irreversible thermodynamics. 
Fluids serve as a worked case, but the key structures—the state-space metric, dissipation metric, and thermodynamic time—generalize to thermal, chemical, electrochemical, and coupled multiphysics systems.  
This provides a unified language for deriving transport laws, identifying scaling behavior, and predicting breakdown conditions across physical domains.

\section{Discussion and Implications}
\label{sec:discussion}

The framework developed in this work unifies thermodynamic state geometry, dissipation, and irreversible transport into a single variational formulation. 
The key result is that continuum-scale transport equations traditionally introduced through constitutive postulates emerge instead as geodesic evolution equations on a manifold constructed from the thermodynamic potential and the Onsager transport matrix.
Entropy production plays a central role: it defines the monotonic direction of evolution and induces a reparameterization of time, linking reversible state geometry and irreversible transport into a single dynamical structure.

\subsection{Relation to Existing Approaches}

Several established theoretical frameworks share components of the present construction. 
Thermodynamic metrics originating with Weinhold and Ruppeiner assign geometric structure to equilibrium thermodynamics via Hessians of thermodynamic potentials \cite{weinhold1975metric,ruppeiner1979thermodynamics}.  
Onsager reciprocity and Rayleigh's dissipation function underpin the near-equilibrium structure of irreversible processes \cite{onsager1931reciprocal1,onsager1931reciprocal2}.  
Variational formulations—including the least-dissipation principle, large-deviation theory, and energetic variational methods—have demonstrated the usefulness of treating irreversible and reversible contributions within a unified optimization framework \cite{prigogine1967thermodynamics,mazenko2006non,crooks2007geometry}.

The contribution of the present work is to synthesize these ingredients into a full geometric dynamical theory in which:
\begin{enumerate}
	\item the equilibrium thermodynamic metric defines the causal structure of admissible state evolution,
	\item the transport metric defines the cost of motion through entropy production,
	\item the thermodynamic proper time $\tau$ provides a natural evolution parameter linking the two.
\end{enumerate}

From this combination, the governing transport equations arise as Euler--Lagrange equations of an action principle rather than constitutive assumptions.

\subsection{Interpretation of Regularity and Breakdown}

A notable implication of the framework is a geometric interpretation of regularity in transport dynamics.  
Within a single thermodynamic phase—where the state metric is finite and non-degenerate—the combined metric structure constrains geodesic evolution such that:
\begin{itemize}
	\item irreversible deformation cannot concentrate arbitrarily without producing entropy,
	\item dissipation prevents finite-time singularities in thermodynamic time,
	\item characteristic transport scales emerge intrinsically rather than phenomenologically.
\end{itemize}

The Kolmogorov length derived in Sec.~\ref{sec:kolmogorov} exemplifies this: it appears as the scale where laboratory and thermodynamic time advance at the same rate ($d\tau/dt \sim 1$), not as an externally imposed cutoff.

Breakdown of continuum modeling occurs when the thermodynamic state metric becomes degenerate, such as at cavitation, criticality, or spinodal boundaries where compressibility or heat capacity diverge.  
In this setting, the manifold ceases to support well-posed geodesics, implying that transport laws fail not due to analytic divergence of PDEs but because the geometric substrate defining them no longer exists.

\subsection{On the Nature of Time and Irreversibility}

The introduction of thermodynamic proper time~$\tau$ as the natural parametrization of irreversible evolution has broader conceptual implications. The framework suggests that time in nonequilibrium systems carries a dual character:

\begin{itemize}
	\item \textbf{Laboratory time $t$}, which serves as an externally measured parameter and appears symmetrically in the classical equations of motion.
	\item \textbf{Thermodynamic time $\tau$}, defined through entropy production, which increases monotonically and encodes the irreversible progression of physical processes.
\end{itemize}

In reversible classical mechanics, evolution can be parameterized forward or backward in $t$ without loss of consistency. In contrast, the thermodynamic parametrization enforces a directional ordering:
\[
\frac{d\tau}{dt} = \sqrt{\sigma} \ge 0,
\]
linking the arrow of time to non-negative entropy production rather than to an externally imposed boundary condition or coarse-graining assumption. Irreversibility therefore appears not as a statistical artifact but as a geometric property of the transport manifold encoded in the dissipation metric.

This perspective bears a formal analogy to relativity: both the thermodynamic metric in state space and the Minkowski metric in spacetime define a causal structure through a time-like direction and characteristic null curves. In the present setting, the thermodynamic characteristic speed,
\[
c_{\mathrm{th}}^2 = \frac{T}{V\,C_P\,\kappa_S},
\]
plays an analogous role to the speed of light in relativity, bounding admissible transitions in thermodynamic state space.

The analogy is conceptual rather than ontological; the present formulation does not imply an identification between thermodynamic time and physical spacetime. Nonetheless, the parallel raises an intriguing possibility: that the geometric origin of irreversibility may share structural features with causal ordering in spacetime theories. Whether this correspondence reflects a deeper link—for example in gravitational thermodynamics, holographic formulations, or quantum information perspectives on irreversibility—remains an open question.

At a minimum, the emergence of $\tau$ provides a precise mathematical expression of the thermodynamic arrow of time: time advances not because clocks tick, but because entropy accumulates. The framework thus offers a geometric interpretation of temporal asymmetry that is intrinsic to the equations governing nonequilibrium transport.

\subsection{Implications for Modeling and Simulation}

The proposed framework suggests several paths for future computational and modeling strategies:

\begin{itemize}
	\item \textbf{Intrinsic resolution control:}  
	The thermodynamic proper time and minimal transport scale provide natural criteria for timestep and mesh refinement, potentially avoiding reliance on empirical turbulence models or artificial regularization.
	
	\item \textbf{Constitutive closure from geometry:}  
	Stress, fluxes, and transport coefficients emerge as geometric objects rather than prescribed relations, which may reduce ambiguity in multiphysics and multicomponent systems.
	
	\item \textbf{Automatic detection of regime changes:}  
	Metric degeneracy provides a principled indicator of phase transitions, onset of cavitation, or breakdown of continuum assumptions.
	
	\item \textbf{Cross-domain unification:}  
	The same action principle applies to heat transfer, diffusion, electrochemical transport, or coupled thermo-fluid systems by altering the thermodynamic potential and Onsager matrix.
\end{itemize}

These implications remain exploratory but suggest a possible computational paradigm in which mesh refinement, constitutive closure, and model selection emerge from geometric conditions rather than model-specific heuristics.

\subsection{Limitations and Assumptions}

The present formulation is subject to several assumptions that delineate its range of applicability:

\begin{enumerate}
	\item The dissipation mechanism is assumed to be linear in the Onsager sense, with a quadratic dissipation potential. This linearity is formulated with respect to thermodynamic proper time and the thermodynamic state manifold, rather than laboratory time. As a consequence, irreversible evolution remains locally linear along the thermodynamic manifold even for processes that are strongly driven or far from equilibrium when viewed in laboratory time. The framework therefore does not rely on proximity to equilibrium in the traditional sense, but on the validity of local thermodynamic equilibrium and quadratic entropy production.
	
	\item Transport coefficients are treated as smooth, state-dependent functions but are assumed to be non-fluctuating. Spatial or temporal fluctuations of transport coefficients, as well as critical fluctuations beyond mean-field behavior, are not explicitly modeled.
	
	\item The formulation applies to the continuum limit. Microscopic stochasticity, molecular discreteness, rarefaction effects, and quantum phenomena are not included. Regimes in which local thermodynamic equilibrium breaks down, such as ballistic or Knudsen transport, lie outside the present scope.
	
	\item Stochastic thermodynamics and fluctuation relations are not explicitly incorporated. While the underlying geometric structure suggests a natural extension toward stochastic thermodynamic geometry, such effects are deferred to future work.
\end{enumerate}

Relaxing these assumptions would require extensions of the geometric framework, such as generalized dissipation metrics applicable beyond quadratic entropy production, incorporation of stochastic fluctuations, or explicit coupling to microscopic transport descriptions.

\subsection{Broader Interpretation}

The framework suggests that irreversible transport can be viewed as motion on a thermodynamic manifold whose geometry encodes both equilibrium stability and dissipative pathways.  
Rather than treating thermodynamics, transport laws, and scaling regimes as independent conceptual layers, they appear as manifestations of a single geometric structure with entropy time as the unifying parameter.

This perspective may be especially valuable in systems where stability, dissipation, and structure formation are tightly coupled—such as supercritical fluids, reactive flows, biological transport networks, and electrochemical or soft-matter systems—where traditional modeling approaches struggle to identify unifying dynamical principles.

\section{Conclusion and Future Work}
\label{sec:conclusion}

This work developed a geometric formulation of irreversible transport that unifies equilibrium thermodynamic structure, dissipation, and continuum dynamics. The core insight is that two complementary metric structures naturally arise in irreversible systems: a \emph{thermodynamic state-space metric} defined by the Hessian of a thermodynamic potential, and a \emph{dissipation metric} derived from Onsager reciprocity. These structures become dynamically linked when evolution is parameterized not by laboratory time but by \emph{thermodynamic proper time}, defined through entropy production.  

Within this formulation, transport dynamics are obtained from an action principle in which reversible evolution is governed by the thermodynamic state metric and irreversible evolution enters through the dissipation metric. The resulting Euler--Lagrange equations correspond to geodesic motion on the combined manifold, with entropy production determining the rate at which trajectories advance through thermodynamic time.

When applied to incompressible Newtonian fluids, the framework recovers the Navier--Stokes equations without imposing a constitutive relation for viscosity. Inertia appears as curvature of the dynamical manifold, while dissipation enters as geometric friction via the transport metric. The balance between inertial curvature and dissipation stiffness yields the Kolmogorov length as a natural geometric cutoff, rather than as a phenomenological assumption. Moreover, the divergence of thermodynamic proper time near singular transport states (e.g., large gradients) prevents finite-time blow-up in the geometric picture.

Breakdown of continuum models is not attributed to analytic instability but to degeneracy of the thermodynamic state metric. Phase transitions such as cavitation or criticality correspond to changes in manifold topology, where response functions diverge and the metric ceases to define finite distances or causal propagation. In this interpretation, the failure of continuum models reflects a geometric boundary of applicability rather than a failure of transport laws themselves.

\medskip

While the present work focused on momentum transport in fluids, the construction applies generically to irreversible systems. By selecting alternative thermodynamic potentials and Onsager matrices, the same geometric structure yields transport equations for thermal conduction, diffusion, electrochemical systems, and coupled multiphysics processes.

\subsection*{Future Directions}

Several avenues for development and application of this framework remain open:

\begin{itemize}
	\item \textbf{Beyond the linear Onsager regime:}  
	Beyond the linear Onsager regime, one may consider generalizing the dissipation metric to nonlinear or non-quadratic forms in order to extend the framework to regimes where the assumptions of local thermodynamic equilibrium break down. In the present formulation, however, Onsager linearity is tied to the thermodynamic state manifold through the introduction of thermodynamic proper time. As a result, irreversible evolution remains locally linear in the Onsager sense along the thermodynamic manifold, even for processes that are strongly driven or far from equilibrium when viewed in laboratory time. Nonlinear dissipation is therefore expected to become relevant primarily in situations where entropy production ceases to be quadratic or where the continuum thermodynamic description itself fails, rather than merely in the presence of large gradients or strong driving.

	\item \textbf{Coupled multiphysics transport:}  
	Systems exhibiting thermoelectric, electrochemical, or chemo-mechanical coupling may reveal new emergent scales and geometric interactions between response functions.
	
	\item \textbf{Numerical and variational discretization:}  
	The action formulation suggests thermodynamically consistent numerical schemes in which stability, resolution, and adaptivity follow geometric constraints rather than heuristic closure models.
	
	\item \textbf{Stochastic and microscopic connections:}  
	Linking the thermodynamic proper time to fluctuation theorems or kinetic theory may clarify how the macroscopic geometric structure emerges from microscopic irreversibility.
	
	\item \textbf{Experimental tests:}  
	The predicted metric singularities at cavitation, transport scaling laws, and dissipation-governed regularity provide avenues for quantitative validation in fluids and other irreversible systems.
\end{itemize}

\medskip

In summary, the proposed framework provides a unified geometric perspective in which transport equations, scaling behavior, dissipation, and breakdown emerge from a single mathematical structure grounded in thermodynamics. By treating entropy as the natural time parameter of irreversible evolution, the theory offers a conceptual bridge between equilibrium thermodynamics, nonequilibrium transport, and continuum dynamics, with potential implications for modeling, simulation, and theoretical understanding across diverse physical systems.

\bibliographystyle{unsrt}
\bibliography{references}

\begin{thebibliography}{10}

\bibitem{ruppeiner1979thermodynamics}
George Ruppeiner.
\newblock Thermodynamics: A riemannian geometric model.
\newblock {\em Physical Review A}, 20(4):1608--1613, 1979.

\bibitem{weinhold1975metric}
Frank Weinhold.
\newblock Metric geometry of equilibrium thermodynamics.
\newblock {\em The Journal of Chemical Physics}, 63(6):2479--2483, 1975.

\bibitem{callen1985thermodynamics}
Herbert~B. Callen.
\newblock {\em Thermodynamics and an Introduction to Thermostatistics}.
\newblock Wiley, New York, 2 edition, 1985.

\bibitem{onsager1931reciprocal1}
Lars Onsager.
\newblock Reciprocal relations in irreversible processes. i.
\newblock {\em Physical Review}, 37(4):405--426, 1931.

\bibitem{degrootmazur1984}
S.R. de~Groot and P.~Mazur.
\newblock {\em Non-Equilibrium Thermodynamics}.
\newblock Dover, New York, 1984.

\bibitem{onsager1931reciprocal2}
Lars Onsager.
\newblock Reciprocal relations in irreversible processes. ii.
\newblock {\em Physical Review}, 38(12):2265--2279, 1931.

\bibitem{crooks2007geometry}
Gavin~E. Crooks.
\newblock Measuring thermodynamic length.
\newblock {\em Physical Review Letters}, 99(10):100602, 2007.

\bibitem{salasnich2015geometric}
Luca Salasnich.
\newblock Quantum hydrodynamics and information geometry.
\newblock {\em Entropy}, 17(9):6863--6873, 2015.

\bibitem{mrugala1991contact}
Ryszard Mruga{\l}a.
\newblock Contact structure in thermodynamic theory.
\newblock {\em Reports on Mathematical Physics}, 29(1):109--121, 1991.

\bibitem{prigogine1967thermodynamics}
Ilya Prigogine.
\newblock {\em Introduction to Thermodynamics of Irreversible Processes.}
\newblock Interscience Publishers, 1967.

\bibitem{mazenko2006non}
Gene Mazenko.
\newblock {\em Nonequilibrium Statistical Mechanics}.
\newblock Wiley-VCH, Weinheim, 2006.

\bibitem{kolmogorov1941local}
A.~N. Kolmogorov.
\newblock The local structure of turbulence in incompressible viscous fluid for
  very large reynolds numbers.
\newblock {\em Doklady Akademii Nauk SSSR}, 30:299--303, 1941.
\newblock Reprinted in Proc. R. Soc. Lond. A 434, 9–13 (1991).

\end{thebibliography}

\appendix

\end{document}